\documentclass[11pt, a4paper]{article} 

\usepackage[margin=1in]{geometry}
\usepackage{setspace}

\usepackage{amsmath, amsfonts, amssymb, amsthm, bm}

\usepackage{graphicx}
\usepackage[hidelinks]{hyperref} 

\theoremstyle{definition}
\newtheorem{property}{Property}
\newtheorem{remark}{Remark}
\theoremstyle{plain}
\newtheorem{theorem}{Theorem}
\newtheorem{lemma}{Lemma}
\newtheorem{proposition}{Proposition}

\begin{document}

\title{On Drift Induced by Local Transition Asymmetry in Combinatorial State Spaces}
\author{Fumio Ishizaki \\
\small Modal Stage Inc., Tokyo, Japan\\
\small \texttt{ishizaki@modalstage.com}}
\date{}
\maketitle

\begin{abstract}

We study stochastic processes on combinatorial state spaces 
with local transition constraints, as arise in local search algorithms. 
We show that asymmetry in local transitions induces a systematic drift 
in a distance process relative to a reference configuration. 
This drift results from the imbalance between inward and outward transitions, 
translating combinatorial multiplicities into directional bias. 
Analyzing the random walk on the Johnson graph, 
we derive explicit expressions for the drift and expected hitting times. 
We also show that locality constraints lead to trajectory-level differences 
that can hinder search trajectories from reaching the target, 
even under identical stationary distributions. 

\end{abstract}

\section{Introduction}

Understanding how the structure of a combinatorial state space
influences search dynamics is a fundamental problem in the analysis
of stochastic optimization algorithms.
In many settings, search processes are constrained to local transitions
on an underlying graph induced by the problem formulation and the
search operator.
While such locality is central to the design of practical algorithms,
it can also introduce systematic biases in the evolution of search
trajectories, 
which may make it difficult for search trajectories to reach the target state,
thereby affecting the efficiency of stochastic optimization algorithms.

A common perspective on optimization difficulty focuses on
properties of individual states, such as the objective landscape
or the distribution of configurations.
For example, approaches based on fitness landscapes characterize
difficulty in terms of ruggedness, local optima, and barriers
defined over objective values (see \cite{Malan2013,Jones1995} and references therein).
Recent work has also moved toward more dynamic characterizations of search processes.
For example, Local Optima Networks (LONs) \cite{Ochoa2008,Ochoa2014,Fieldsend2025} 
represent local optima and their connectivity,
while Search Trajectory Networks (STNs) \cite{Ochoa2021,Thomson2024} 
represent observed search trajectories as graphs.
While these approaches provide valuable descriptive insights into search behavior,
they do not explicitly identify the mechanisms 
by which local transition structure generates 
systematic trajectory-level bias. 

Related perspectives arise in statistical physics, where the
interplay between energy and entropy is captured through free
energy landscapes (see \cite{Mezard2009,Nishimori2001} and references therein).
These approaches provide a powerful description of equilibrium
behavior, in which the probability of observing a configuration
is determined by its free energy.
While entropy plays a central role, it contributes primarily
through the weighting of states, thereby giving rise to entropic
barriers and free energy differences defined over states or
equilibrium distributions.

A related line of work is found in the study of Markov Chain
Monte Carlo (MCMC) (e.g., \cite{RobertCasella2004}), where stochastic
processes are analyzed through their stationary distributions
and the rate at which these distributions are approached.
In this framework, performance is characterized by mixing
properties, such as spectral gaps or mixing times, which
quantify convergence to equilibrium \cite{LevinPeresWilmer2009}.

While these perspectives provide useful insights, they primarily
focus on static properties of states, their connectivity, 
or stationary distributions over states. 
However, search algorithms operate as stochastic processes evolving over time,
and optimization performance is inherently influenced by transient behavior. 
Accordingly, the performance of search algorithms should be measured 
not by the probability of the target under a stationary distribution, 
but by the ability of the search process to generate trajectories 
that reach the target within a finite time. 
This gap motivates a shift from state-based descriptions
to trajectory-based analysis.

Another important line of work studies search processes through
drift analysis (see \cite{Lengler2020} and references therein), which characterizes the expected progress of a
stochastic process toward a target configuration.
This approach has been widely used in the analysis of randomized
search heuristics (e.g., \cite{HoosStutzle2005,AugerDoerr2011}) 
and evolutionary algorithms (e.g., \cite{He2001,NeumannWitt2010,DoerrNeumann2020}).
While these perspectives provide useful insights, they primarily
focus either on static properties of the state space or on
aggregate measures of progress.
They do not explicitly describe how local transition structure
translates combinatorial multiplicities into a systematic bias
in trajectory evolution.
In particular, they do not explicitly distinguish between the static
distribution of configurations and the dynamical mechanism
by which local transition asymmetry translates this
distribution into a systematic drift.

In this paper, we take a complementary perspective and analyze
search dynamics through the drift of a distance process defined
with respect to a reference configuration.
To analyze the mechanism inducing the search dynamics, 
it is useful to decompose the drift
of the distance process into two components:
one arising from variation in the objective function and another
arising from asymmetry in local transitions induced by the underlying graph structure, 
which is determined by the combination of the problem structure and the search operator. 
These two components correspond, respectively, to
objective-induced effects and structure-induced effects
in the evolution of search trajectories.
This decomposition separates the contribution of the objective
from that of the transition structure, allowing the structural
component of the dynamics to be examined independently.
Hence, it is particularly useful for understanding how
structural constraints of the state space influence trajectory evolution,
independently of objective-based effects.

A key point of this work is to clarify the mechanism that links
combinatorial structure to trajectory-level dynamics.
While it is well known that combinatorial state spaces exhibit
strong multiplicity effects, such multiplicity alone does not
induce a systematic drift.
For instance, under independent sampling, a distribution
over configurations induced by combinatorial multiplicities
arises without any directional bias. 
The crucial mechanism instead lies in the interaction between
combinatorial multiplicities and local transition constraints.
When transitions are restricted by the underlying graph structure,
the imbalance in the number of accessible neighboring states
induces an asymmetry between outward and inward moves.
This asymmetry provides a mechanism that translates the static
combinatorial structure into a systematic drift in trajectory dynamics. 
This distinction between static multiplicity and dynamically
induced drift is central to the trajectory-level perspective
developed in this paper.

This mechanism is reminiscent of entropy-driven effects
in statistical physics, where gradients of entropy
can induce effective forces on macroscopic variables.
However, the present formulation differs in that the drift
arises directly from asymmetry in local transition
probabilities on the underlying graph, rather than from
a gradient of a free-energy functional.
Related viewpoints also appear in stochastic thermodynamics (see \cite{Peliti2024} and references therein),
where entropy production governs the evolution of 
probability distributions, although the present work
focuses on drift at the level of a reduced coordinate.

To make the mechanism yielding the structure-induced drift explicit, 
we study the uniform random walk
on the Johnson graph \cite{GodsilRoyle2001}, which provides a canonical model of local
search dynamics for $k$-of-$n$ subset selection problems. 
The $k$-of-$n$ subset selection problem arises in a wide range of applications 
including feature selection, combinatorial design, and portfolio selection 
(see, e.g., \cite{Guyon2003,BlumLangley1997,Benidis2018,Sakurai2024} and references therein). 
Owing to the strong symmetry properties of this graph, the dynamics
can be reduced to a one-dimensional distance process, allowing an
explicit characterization of the drift induced by local transitions.
This setting isolates the structural component of the dynamics
by removing the influence of the objective function.

Our analysis shows that the imbalance between inward and outward
local transitions induces a systematic bias in the distance process.
This bias can be expressed explicitly as a function of the distance
from the reference configuration and leads to a mean-reverting
behavior toward an equilibrium distance determined by the transition
structure. 
We further show how this drift is related to, but distinct from,
the entropy landscape defined by the combinatorial distribution
of states.

The Johnson graph provides a tractable setting in which this
mechanism can be analyzed in closed form.
For more general graphs, the distance process need not be Markovian,
and the drift may depend on the specific structure of the underlying
graph.
Understanding how structural asymmetry in local transitions shapes
search dynamics in such settings remains an important direction
for future work.

The main contributions of this paper are as follows.
First, we introduce a general framework that decomposes
the drift of a distance process into objective-driven
and structure-induced components.
Second, we show that, even in the absence of objective variation,
asymmetry in local transitions induces a systematic drift
that governs trajectory evolution.
Third, we provide a closed-form analysis of this mechanism
on the Johnson graph, including explicit expressions for
the drift and expected hitting times.

The remainder of the paper is organized as follows.
Section~2 introduces a general framework for analyzing drift
induced by local transitions.
Section~3 describes the structure of the Johnson graph.
Section~4 analyzes the combinatorial structure of distance shells,
and Section~5 introduces the corresponding entropy landscape.
Section~6 characterizes the induced search dynamics, including
explicit expressions for the drift and expected hitting times.
Section~7 presents numerical experiments illustrating the
trajectory behavior.
Section~8 discusses the implications of the results,
and Section~9 concludes the paper with a summary
and directions for future research.

\section{A General Framework for Entropy-Driven Drift}

We begin by describing a general framework for analyzing the trajectory dynamics of stochastic search processes 
on combinatorial state spaces.

We represent the state space of a combinatorial optimization problem as a graph,
where vertices correspond to feasible configurations and edges correspond to possible local transitions
between configurations.
The graph structure is determined by the problem structure and the local transition rules
of the search operator \cite{Jones1995}.
We refer to this graph as the \emph{underlying graph}.
In this framework, the search process can be viewed as a stochastic process on the underlying graph.

The state space admits three equivalent perspectives:
as configurations of a combinatorial optimization problem,
as states of a stochastic process,
and as vertices of the underlying graph.
We use the terms \emph{configuration}, \emph{state}, and \emph{vertex}
interchangeably throughout the paper.

In this work, we distinguish two components of drift that influence trajectory evolution.
The first arises from variation in the objective function across states.
The second arises from asymmetry in local transitions induced by the underlying graph structure.
We call these contributions \emph{objective-driven drift} and
\emph{structure-induced drift}, respectively.

To isolate the effect of the underlying graph structure,
we consider the uniform random walk on the underlying graph as a baseline process.
This baseline removes any explicit objective-based bias and reveals
the drift induced purely by the transition structure.
In other words, the drift is entirely structural and reflects 
asymmetry in local transitions, independent of the objective function. 

Let $G = (\Omega,E)$ be the underlying graph, and let $\{X_t\}_{t = 0}^{\infty}$
be the random walk on $G$.
Fix a reference configuration $x^* \in \Omega$,
which may represent a target or optimal state.
We define the distance $v(x,x^*)$ as the shortest-path distance
from $x$ to $x^*$ on the underlying graph.

The evolution of the process can be characterized through the local drift
\begin{equation}
\Delta(x)
=
\mathbb{E}\left[v(X_{t+1},x^*) - v(X_t,x^*) \mid X_t = x \right].  
\label{eq:drift-general}
\end{equation}
In general, this drift (\ref{eq:drift-general}) depends on the specific state $x$,
and not only on its distance from the target.
Consequently, the distance process $\{Y_t\}$ defined by $Y_t = v(X_t,x^*)$ $(t=0,1,2,\ldots)$ is not, in general,
a Markov process, as the transition probabilities depend on the full state $x$
and not only on its distance from the reference configuration.

To express the drift in terms of local transitions on the underlying graph, 
it is useful to classify neighboring states according to 
how they change the distance to the reference configuration.
More precisely, for $x \in \Omega$ and $l \in \mathbb{Z}$, define
\begin{equation}
\Gamma(x, l) =
\left\{ x' : v(x',x^*) - v(x,x^*) = l,\ (x,x') \in E \right\}.
\label{eq:Gamma}
\end{equation}
By the definition of the graph distance,
the distance can change by at most one along an edge.
Therefore, the only possible values of $l$ are $-1,0,1$,
and $\Gamma(x,l)=\emptyset$ for $l\le -2$ or $l\ge 2$.

Under the uniform random walk, transitions occur uniformly over neighbors of $x$.
Therefore, the drift (\ref{eq:drift-general}) can be expressed in terms of the relative number of
inward and outward transitions, and is determined by their imbalance, 
which can be represented in terms of (\ref{eq:Gamma}): 
\begin{equation}
\Delta(x)
=
\frac{-|\Gamma(x,-1)| + |\Gamma(x,1)|}
{\sum_{l=-1}^{1}|\Gamma(x,l)|}.
\label{eq:EDD-general}
\end{equation}
This expression (\ref{eq:EDD-general}) makes explicit how asymmetry in local transitions
induces a drift in the distance process.
We refer to this structural component of the drift as
\emph{entropy-driven drift}, 
reflecting its connection to the entropy landscape introduced later.
More generally, for arbitrary search processes,
the observed drift may be decomposed into 
objective-driven and entropy-driven components,
where the latter is characterized through the transition asymmetry
of the baseline uniform random walk.

\section{Structure of the Underlying Graph}

To make the mechanism of entropy-driven drift explicit,
we consider the $k$-of-$n$ subset selection problem with a random swap operator,
which provides a canonical setting in which the drift can be analyzed in closed form.

In this section, we introduce the underlying graph 
for $k$-of-$n$ subset selection problem with random swap operator. 
The structural organization of the underlying graph will serve 
as the foundation
for analyzing the induced search dynamics in subsequent sections. 

We consider subset selection problems in which exactly $k$ elements are chosen
from a ground set of size $n$. 
Throughout the paper, we assume that $1 \leq k \leq n-1$, excluding the trivial cases of selecting no elements or all elements. 

We first consider the feasible space $\Omega$ of the problem, i.e., 
the vertex set of the underlying graph.
Let $[n] = \{1,2,\dots,n\}$ denote the ground set of $n$ elements. The feasible space consists of all $k$-element subsets of $[n]$:  
\begin{equation}
\Omega = \binom{[n]}{k}
= \{ x \subseteq [n] : |x| = k \}, 
\end{equation}
where $\binom{[n]}{k}$ denotes the family of all $k$-subsets of $[n]$.
Each configuration corresponds to a $k$-subset $x \in \Omega$.

Next we consider the edge set $E$ of the underlying graph, 
which is determined by the one-step transition of the search process.
Two configurations $x,y \in \Omega$ are adjacent if they differ by a
single swap, that is,
\begin{equation}
|x \setminus y| = |y \setminus x| = 1, 
\end{equation}
where $x \setminus y$ denotes the set difference. 
$(x, y) \in E$ if and only if $x$ and $y$ are adjacent in this sense. 

The resulting underlying graph $G = (\Omega, E)$ is known as the Johnson graph 
$J(n,k)$ \cite{GodsilRoyle2001}.
Each vertex in
the Johnson graph $J(n,k)$ has degree $k(n-k)$.

We now introduce a distance function $v$ on the Johnson graph, 
which will be used to analyze the trajectory dynamics of the search process.
The distance between two configurations $x$ and $y$ is defined as the number of
swaps required to transform one subset into the other:
\begin{equation}
v(x,y) = |x \setminus y| = |y \setminus x| = \frac{|x \triangle y|}{2},
\label{eq:distance}
\end{equation}
where $x \triangle y = (x \setminus y) \cup (y \setminus x)$ denotes the
symmetric difference between the two sets.
This distance $v$ defines the natural graph metric on the Johnson graph, 
and it corresponds to the shortest-path distance under the swap operation.
This definition (\ref{eq:distance}) of the distance function is consistent with the one introduced in the general framework of Section~2,

In the theoretical analysis that follows, we consider a reference state on the feasible space, 
i.e., a reference vertex in the Johnson graph.
Throughout the paper, we fix a reference state, which may represent an optimal solution.
If multiple optimal solutions exist, we select one arbitrarily 
as the reference state.
Note that because the Johnson graph is vertex-transitive \cite{GodsilRoyle2001} (see Property~1 below), 
the choice of reference vertex does not affect 
the structure of the analysis itself. 

Let $x^*$ denote a fixed reference state, which may represent an
optimal solution.
For each integer $d \in \{ 0, 1, \dots, d_{\max} \}$ where $d_{\max} = \min(k,n-k)$, we define the \emph{distance shell}
\begin{equation}
\Omega_d =
\{ x \in \Omega : v(x,x^*) = d \}.
\end{equation}
These shells form a partition of the feasible space:
\begin{equation}
\Omega = \bigcup_{d=0}^{d_{\max}} \Omega_d. 
\end{equation}
This shell decomposition provides a natural coordinate system for 
the feasible space relative to the reference state $x^*$.

The Johnson graph exhibits strong symmetry properties,
which play a central role in simplifying the analysis of the induced dynamics.
These properties are summarized as follows.
\begin{property}[Symmetry of the Johnson Graph]
The Johnson graph $J(n,k)$ satisfies 
the following transitivity properties \cite[Chapter 4]{GodsilRoyle2001}:
\begin{enumerate}
\item \textbf{Vertex-transitivity}: The automorphism group of $J(n,k)$ acts transitively on the set of vertices $\Omega$. 
This ensures that global structural properties of the state space are invariant 
across configurations.
\item \textbf{Distance-transitivity}: 
The graph is distance-transitive, meaning that any pair of vertices at distance $d$ can be mapped 
to any other pair at the same distance by an automorphism. 
This property, which is stronger than vertex-transitivity, implies that for any fixed target $x^* \in \Omega$, 
all vertices at the same distance from $x^*$ are structurally and probabilistically equivalent.

\end{enumerate}
\end{property}

\section{Distance Shell Structure}

In this section, we analyze the combinatorial structure of the distance shells
introduced in the previous section.
This structure determines how the volume of the state space is distributed
as a function of distance from the reference state, and provides the
basis for understanding the induced search dynamics in subsequent sections.

We first consider the cardinalities $|\Omega_d|$ $(d = 0, 1, \dots, d_{\max})$ of the distance shells $\Omega_d$.
The following proposition provides an explicit formula for the number of states in each distance shell, 
i.e., for the cardinalities $|\Omega_d|$. 
\begin{proposition}
Let $x^*$ be a fixed reference state.
The number of states at distance $d$ $(d = 0, 1, \dots, d_{\max})$ from $x^*$ is
\begin{equation}
|\Omega_d|
=
\binom{k}{d}\binom{n-k}{d}.
\label{eq:shell_size}
\end{equation}
\end{proposition}
\begin{proof}
To obtain a state at distance $d$ from $x^*$, we must remove
$d$ elements from $x^*$ and replace them with $d$ elements from
$[n]\setminus x^*$.
The number of ways to remove $d$ elements from $x^*$ is
$\binom{k}{d}$, and the number of ways to choose $d$ replacement
elements from the $n-k$ elements outside $x^*$ is
$\binom{n-k}{d}$.
Multiplying these choices gives the result.
\end{proof}

We next analyze how the shell cardinalities vary with distance,
as this variation plays an important role in the subsequent analysis of
local transition behavior. 
The following proposition shows that the shell cardinalities form a unimodal sequence as a function of 
the distance $d$ from the reference state $x^*$.
\begin{proposition}
The sequence $\{|\Omega_d|\}_{d=0}^{d_{\max}}$
is unimodal.
\end{proposition}
\begin{proof}
From Eq.~(\ref{eq:shell_size}), 
for $d=0,1,\ldots,d_{\max}-1$, 
the ratio of the cardinalities of adjacent shells is given by 
\begin{align}
\frac{|\Omega_{d+1}|}{|\Omega_d|}
&=
\frac{\binom{k}{d+1}\binom{n-k}{d+1}}{\binom{k}{d}\binom{n-k}{d}} \nonumber \\
&=
\frac{(k-d)(n-k-d)}{(d+1)^2}.
\label{eq:ratio_adjacent_shells}
\end{align}
We then consider 
\begin{align} 
\frac{|\Omega_{d+1}|}{|\Omega_d|} - 1 
&= 
\frac{(k-d)(n-k-d)}{(d+1)^2} - 1  \nonumber \\
&=
\frac{-(n+2)d + k(n-k)-1}{(d+1)^2}. 
\label{eq:ratio_minus_one}
\end{align}
Since the denominator $(d+1)^{2} >0$ for all $d=0,\ldots,d_{\max}-1$,
the sign of $|\Omega_{d+1}|/|\Omega_d| -1$
is determined by the numerator, which is a 
linear function of $d$ given by 
\begin{equation}
-(n+2)d + k(n-k)-1.
\label{eq:linear}
\end{equation}
Hence the sign of $|\Omega_{d+1}|/|\Omega_d| -1$ can change from positive to negative at most once as $d$
increases.
Therefore the sequence $\{|\Omega_d|\}$ first increases
and then decreases as $d$ increases, and is unimodal. 
\end{proof}

The shell cardinalities admit a natural probabilistic interpretation 
associated with $k$-of-$n$ subset selection problems.
Consider drawing $k$ elements without replacement from a population
of size $n$ containing $k$ distinguished elements.
Let $Z$ denote the number of distinguished elements that are selected.
Then $Z$ follows a hypergeometric distribution 
and the probability of $Z=k-d$ is given by 
\begin{equation}
\mathbb{P}(Z=k-d)
=
\frac{\binom{k}{k-d}\binom{n-k}{n-k-d}}{\binom{n}{k}} 
= 
\frac{\binom{k}{d}\binom{n-k}{d}}{\binom{n}{k}}.
\label{eq:hypergeometric}
\end{equation}
Up to normalization, the shell cardinalities given by Eq.~(\ref{eq:shell_size}) coincide 
with the hypergeometric distribution given by Eq.~(\ref{eq:hypergeometric}). 
Since the hypergeometric distribution is unimodal, 
the shell cardinalities increase for small $d$,
reach a maximum at an intermediate distance,
and then decrease.

We finally consider the maximum shell cardinality and the corresponding distance from the reference state.
The following proposition shows that the shell cardinality is maximized away from the reference state.
\begin{proposition}
Assume $n \ge 3$.
The shell cardinality $|\Omega_d|$ is maximized at an integer
\begin{equation}
d \in \{\lfloor \hat d^* \rfloor,\ \lceil \hat d^* \rceil\}, 
\end{equation}
where
\begin{equation}
\hat d^*=\frac{k(n-k)-1}{n+2}.
\label{eq:dstar}
\end{equation}
In particular, the maximizer satisfies $d \ge 1$.
\end{proposition}
\begin{proof}
From the unimodality of the shell cardinalities shown in Proposition~2,
the shell cardinality is maximized at an integer $d$ where
\begin{equation}
\frac{|\Omega_d|}{|\Omega_{d-1}|} \ge 1
\quad\text{and}\quad
\frac{|\Omega_{d+1}|}{|\Omega_d|} \le 1
\label{eq:conditions}
\end{equation}
hold, with the obvious modification at the boundary.

We first consider the case $d \ge 1$. 
For $d\ge 1$, using Eq.~(\ref{eq:ratio_minus_one}) from the proof of Proposition~2,
the conditions expressed in Eqs.~(\ref{eq:conditions}) are equivalent to
\begin{equation}
-(n+2)(d-1)+k(n-k)-1 \ge 0
\label{eq:condition1}
\end{equation}
and
\begin{equation}
-(n+2)d+k(n-k)-1 \le 0.
\label{eq:condition2}
\end{equation}
Hence, from Eqs.~(\ref{eq:condition1}) and (\ref{eq:condition2}), we have 
\begin{equation}
\hat d^* \le d \le \hat d^*+1,
\qquad
\hat d^*=\frac{k(n-k)-1}{n+2}.
\end{equation}
Therefore, any maximizer belongs to
$\{\lfloor \hat d^* \rfloor,\ \lceil \hat d^* \rceil\}$.

We then treat the boundary case $d=0$. 
From Eq.~(\ref{eq:shell_size}), since 
\begin{equation}
|\Omega_0|=1
\quad\text{and}\quad
|\Omega_1|=\binom{k}{1}\binom{n-k}{1}=k(n-k),
\end{equation}
we have
\begin{equation}
\frac{|\Omega_1|}{|\Omega_0|}=k(n-k).
\end{equation}
Under the assumptions $1\le k\le n-1$ and $n\ge 3$, we have
\begin{equation}
k(n-k)\ge 2.
\end{equation}
Thus
\begin{equation}
|\Omega_1|>|\Omega_0|,
\end{equation}
so the shell cardinality cannot be maximized at $d=0$.
Therefore, the maximizer satisfies $d\ge 1$.
\end{proof}

While the above result holds under the minimal assumption $1 \le k \le n-1$,
in typical subset selection settings where $2 \le k \le n-2$,
the maximizer lies at an intermediate distance from the reference state.
In particular, in typical settings with $1 \ll k \ll n$ in $k$-of-$n$ subset selection problems, 
the maximizing distance lies well away from the boundary,
indicating that most configurations are concentrated at intermediate distances.

Recall that the Johnson graph is vertex-transitive, so the shell structure revealed in this section  
is the same relative to any reference state.

\section{Entropy Landscape}

In this section, we introduce the entropy landscape over the feasible space
defined by the shell cardinalities derived in the previous section.
This provides a static description of how combinatorial volume is distributed
as a function of distance from the reference state.

Let $S(d) = \log |\Omega_d|$ 
denote the entropy associated with the distance shell
$\Omega_d$.
Using (\ref{eq:shell_size}), we obtain
\begin{equation}
S(d)
=
\log \binom{k}{d}
+
\log \binom{n-k}{d}.
\label{eq:entropy-landscape}
\end{equation}
The entropy landscape is the function $S(d)$ viewed over the distance coordinate $d$.
It is therefore entirely determined by the shell cardinalities of the Johnson graph.

\subsection{Entropy Increment}

Since the logarithm is strictly increasing, the entropy is maximized
at the same distance where the shell cardinality is maximized.
From Proposition~3, 
the shell cardinality $|\Omega_d|$ attains its
maximum at an integer $d \in \{\lfloor \hat{d}^* \rfloor, \lceil \hat{d}^* \rceil\}$, and 
thus the entropy attains its maximum at the same distance $d$, too. 
Importantly, this distance is typically separated from the reference
configuration, especially in high-dimensional settings.

To understand the shape of the entropy landscape, we examine
the entropy increment defined as 
\begin{equation}
\Delta S(d)
=
S(d+1)-S(d).
\label{eq:entropy-increment-def}
\end{equation}
Using the ratio (\ref{eq:ratio_adjacent_shells}) of the cardinalities of adjacent shells,  
we obtain
\begin{equation}
\Delta S(d)
=
\log \frac{(k-d)(n-k-d)}{(d+1)^2}.
\label{eq:entropy-increment}
\end{equation}
From the unimodality of the shell cardinality shown  in Proposition~2 and the monotonicity of the logarithm,  
we see that 
\begin{equation}
\Delta S(d)>0 \quad \text{for}\quad d \le \lfloor \hat{d}^* \rfloor,
\end{equation}
and
\begin{equation}
\Delta S(d)<0 \quad \text{for}\quad d \ge \lceil \hat{d}^* \rceil, 
\end{equation}
where $\hat{d}^*$ is given by Eq.~(\ref{eq:dstar}). 
As $d$ increases, 
the entropy thus increases as the distance approaches the
entropy-maximizing shell and decreases beyond it.

\subsection{Distribution over Distance Shells}

The shell cardinalities also admit a natural probabilistic interpretation.
If a configuration $X$ is drawn uniformly at random from the feasible space $\Omega$,
the probability that $v(X,x^*)=d$ is given by $|\Omega_d|/|\Omega|$. 
This interpretation is consistent with the hypergeometric
structure of the shell cardinalities discussed in Section~4. 
Thus, the entropy landscape $S(d)=\log|\Omega_d|$ describes how the uniform measure
over $\Omega$ is distributed across distances from the reference state.

From this viewpoint, 
the induced distribution over $d$ is maximized at
distances near $\hat{d}^*$, namely at
$d \in \{\lfloor \hat{d}^* \rfloor, \lceil \hat{d}^* \rceil\}$. 
Hence, shells near $\hat{d}^*$ carry the largest combinatorial weight and are
the most typical under the uniform measure.
This shows that the feasible space is combinatorially biased toward intermediate
distances from the reference state.
In particular, configurations close to or far from the reference state are
relatively less frequent compared to those near $\hat{d}^*$.

This probabilistic viewpoint provides a natural interpretation of the entropy landscape:
the function $S(d)$ captures how the combinatorial volume of the feasible space
is distributed across distances.
This structural bias plays a central role in shaping the
local transition dynamics analyzed in the next section. 

\subsection{Free-Energy Perspective}

The entropy landscape derived above also admits an interpretation
in terms of free energy (see \cite{Mezard2009,Nishimori2001} and references therein).
In stochastic search processes, the dynamics can be interpreted
through the lens of the Helmholtz free energy
\begin{equation}
F = E - T S,
\end{equation}
where $E$ denotes an objective function, $S$ is the entropy,
and $T$ is a parameter controlling the relative influence of
energy and entropy.

In the present setting, the entropy term $S(d)$ is determined
solely by the combinatorial structure of the feasible space,
independently of the objective function.
This highlights a fundamental distinction between two sources
of bias in search dynamics:
objective-driven effects through $E$, and 
combinatorial-structure-induced effects through $S$, 
which are typically combined in standard
free-energy formulations.
As illustrated in the experimental results (Section~7), 
the interaction between 
these two components influences the behavior of search trajectories.

\section{Search Dynamics with Entropy-Driven Drift}

In this section, we specialize the general framework of Section~2
to the Johnson graph.
The uniform random walk on this graph serves as a baseline process
that isolates the structural component of the drift.
Owing to the distance-transitivity of the Johnson graph,
the local drift defined in Section~2 can be expressed solely as a function
of the distance from the reference configuration.
This allows us to obtain a closed-form characterization of the
entropy-driven drift at the trajectory level.

\subsection{Random Walk on the Johnson Graph}

We consider the uniform random walk $\{ X_t \}_{t=0}^{\infty}$ on the Johnson graph 
$J(n,k)$, which serves as a canonical model of local search dynamics for $k$-of-$n$ subset selection problems.
From a configuration $X_t \in\Omega=\binom{[n]}{k}$ at time $t$ ($t=0,1,2,\ldots$), 
a step of
the walk replaces one element of $X_t$ with one element from 
$[n]\setminus X_t$, where both elements are chosen uniformly at random as $X_{t+1}$. 
This random walk is a Markov chain on the finite state space $\Omega$. 
We first summarize basic structural properties of this random walk,
which will be used in the analysis of the induced distance process.

\begin{lemma}
The random walk $\{ X_t \}$ on the Johnson graph $J(n,k)$ is a
finite-state Markov chain that is irreducible and aperiodic.
Consequently, it admits a unique stationary distribution 
and its stationary distribution is uniform on the feasible space $\Omega$. 
\end{lemma}
\begin{proof}
Irreducibility follows from the connectivity of the Johnson
graph: any $k$-subset can be transformed into any other by a
finite sequence of swaps.

Since the Johnson graph is vertex-transitive (Property~1), 
the transition probabilities are symmetric. 
Thus, to prove aperiodicity, it suffices to show that the graph
contains an odd cycle. 
If $k=1$, then $J(n,1)=K_n$, where $K_n$ is the complete graph on $n$ vertices. 
Thus, it contains triangles for
$n\ge3$.
If $k\ge2$, choose distinct $a,b\in X$ and $c\notin X$, where $X$ denotes 
any configuration $X \in \Omega$. 
Then the configurations $X$, $X\setminus\{a\}\cup\{c\}$, and $X\setminus\{b\}\cup\{c\}$ 
form a triangle in $J(n,k)$.
Hence the graph is non-bipartite and the Markov chain is aperiodic.

Finally, as $J(n,k)$ is vertex-transitive (Property~1), 
it is a regular graph \cite{GodsilRoyle2001}. 
For any random walk on a connected, undirected regular graph, 
the stationary distribution is unique and uniform on $\Omega$.
\end{proof}

\subsection{Distance Process}

To analyze the search dynamics, we consider the distance between the current state $X_t$ and 
a fixed reference configuration $x^* \in \Omega$. 
For $t=0,1,2,\ldots$, let 
\begin{equation}
Y_t = v(X_t, x^*)
\label{eq:distance_process}
\end{equation}
be the distance process,  representing the shortest-path distance from $X_t$ 
to the fixed reference configuration $x^*$ on the Johnson graph.

\begin{proposition}
The distance process $\{ Y_t \}$ determined by Eq.~(\ref{eq:distance_process}) is 
a Markov chain on $\{0, 1, \dots, d_{\max}\}$. 
More specifically, $\{ Y_t \}$ is a birth-and-death chain on $\{0, 1, \dots, d_{\max}\}$. 
\end{proposition}
\begin{proof}
Since each step in the swap random walk $\{ X_t \}$ changes the distance by at most one, 
the distance $Y_t$ can change by at most one in a single transition. 
By distance-transitivity of the Johnson graph $J(n,k)$ (Property~1), 
for each $d=0,1,\ldots,d_{\max}$ and each $d' \in \{d-1,d,d+1\}$, 
the transition probability
\begin{equation}
\mathbb{P}(X_{t+1}\in \Omega_{d'} \mid X_t=x)
\end{equation}
is constant over all $x \in \Omega_d$. 
This implies that the transition probabilities depend only on the distance $d$, 
not on the specific state $x \in \Omega_d$, satisfying the condition for lumpability \cite{KemenySnell1976}. 
The lumpability condition ensures that the aggregated process $\{ Y_t \}$ retains the Markov property. 

Consequently, the distance process $\{Y_t \}$ is a birth-and-death chain \cite{LevinPeresWilmer2009} 
on the state space $\{0, 1, \dots, d_{\max}\}$. 
\end{proof}

We now examine the transition probabilities of the distance process $\{ Y_t \}$. 
For $d=0,1,\ldots, d_{\max}$, let $p_d$ denote the probability of moving from distance $d$ to $d+1$,
$q_d$ denote the probability of moving from distance $d$ to $d-1$,
and $r_d$ denote the probability of remaining at distance $d$. 
We now compute $|\Gamma(x, l)|$ where $\Gamma(x, l)$ is defined by Eq.~(\ref{eq:Gamma}). 
Due to the distance-transitivity of $J(n,k)$ (Property~1), 
for any $x_d \in \Omega_d$ $(d=0,1,\ldots,d_{\max})$, 
we obtain 
\begin{equation}
|\Gamma(x_d, 1)| = (k-d)(n-k-d), 
\label{eq:Gamma1} 
\end{equation}
\begin{equation}
|\Gamma(x_d, -1)| = d^2, 
\label{eq:Gamma-1} 
\end{equation}
\begin{equation}
|\Gamma(x_d, 0)| = d(n-k-d) + d(k-d) = -2 d^{2} + nd.  
\label{eq:Gamma0}
\end{equation}
From Eqs.~(\ref{eq:Gamma1}), (\ref{eq:Gamma-1}), and (\ref{eq:Gamma0}), 
the transition probabilities of the distance process $\{Y_t\}$ are then expressed as follows:
\begin{align}
q_d &= \displaystyle{\frac{|\Gamma(x_d, -1)|}{\sum_{l=-1}^{1} |\Gamma(x_d, l)|}} = \displaystyle{\frac{d^2}{k(n-k)}},  \label{eq:qd} \\
p_d &= \displaystyle{\frac{|\Gamma(x_d, 1)|}{\sum_{l=-1}^{1} |\Gamma(x_d, l)|}} = \displaystyle{\frac{(k-d)(n-k-d)}{k(n-k)}},  \label{eq:pd} \\
r_d &= \displaystyle{\frac{|\Gamma(x_d, 0)|}{\sum_{l=-1}^{1} |\Gamma(x_d, l)|}} = \displaystyle{\frac{d(n-2d)}{k(n-k)}}. 
\label{eq:rd}  
\end{align}
Since $d\le d_{\max}=\min(k,n-k)$, these probabilities are nonnegative.
The expressions (\ref{eq:qd}), (\ref{eq:pd}), and (\ref{eq:rd}) naturally incorporate the boundary conditions:
\begin{itemize}
\item At the target state $d=0$, we have $q_0 = 0$, reflecting that the process cannot move to a negative distance.
\item At the maximum distance $d=d_{\max}$, we have $p_{d_{\max}} = 0$, ensuring the process remains within the feasible state space.
\end{itemize}
This confirms that the distance process $\{Y_t\}$ is a well-defined birth-and-death chain 
on the finite state space $\{0, 1, \dots, d_{\max}\}$.

\subsection{Entropy-Driven Drift on the Johnson Graph}

We now specialize the entropy-driven drift $\Delta(x)$ defined in
Eq.~(\ref{eq:EDD-general}) to the Johnson graph.
Because of the distance-transitivity of the Johnson graph (Property~1), the drift depends only on
the distance $d$ from the reference configuration $x^*$.
This corresponds to the structural drift induced by the 
baseline uniform random walk introduced in Section~2.
For notational simplicity, we write
\begin{equation}
\mathbb{E}[\Delta d \mid d]
=
\mathbb{E}[Y_{t+1}-Y_t \mid Y_t=d] 
= 
\mathbb{E}[v(X_{t+1}, x^*)-v(X_{t}, x^*) \mid X_t=x] 
\end{equation}
for any $x \in \Omega_d$ $(d=0,1,\ldots,d_{\max})$.

Using the expressions (\ref{eq:qd}) and (\ref{eq:pd}) for the 
transition probabilities of the distance process $\{ Y_t \}$ in 
Eq.~(\ref{eq:EDD-general}), 
we can express the entropy-driven drift as
\begin{align}
\mathbb{E}[\Delta d \mid d]
&= p_d - q_d \label{eq:pd-qd} \\ 
&= \frac{(k-d)(n-k-d)-d^2}{k(n-k)} \nonumber \\
&= 1 - \frac{nd}{k(n-k)}. \label{eq:drift}
\end{align}

\paragraph{Connection to the Entropy Increment.}

The entropy landscape and the transition probabilities are related
through the detailed balance identity 
for $d=0,\ldots,d_{\max}-1$, 
\begin{equation}
|\Omega_d|p_d = |\Omega_{d+1}|q_{d+1}.
\end{equation}
Equivalently,
\begin{equation}
\frac{p_d}{q_{d+1}}
=
\frac{|\Omega_{d+1}|}{|\Omega_d|}
=
\exp(S(d+1)-S(d)).
\label{eq:pd-qd-entropy}
\end{equation}
Equation (\ref{eq:pd-qd-entropy}) provides a direct link between the static entropy
landscape and the reversible structure of the distance process. 
However, as shown in Eq.~(\ref{eq:pd-qd}), the local drift at distance \(d\) is determined by
\(p_d-q_d\), or equivalently by the imbalance between outward and
inward transitions from the same shell.
Thus the entropy increment and the entropy-driven drift are closely
related but not identical quantities.

From Eq.~(\ref{eq:drift}),
we see that the entropy-driven drift vanishes at 
\begin{equation}
d^*=\frac{k(n-k)}{n}, 
\label{eq:dstar_drift}
\end{equation}
which can be interpreted as the stochastic equilibrium point of the search process. 
We can rewrite Eq.~(\ref{eq:drift}) 
in the linear form of the stochastic equilibrium point $d^*$ given by Eq.~(\ref{eq:dstar_drift}) as
\begin{equation}
\mathbb{E}[\Delta d \mid d]
=
-\frac{n}{k(n-k)}(d-d^*).
\label{eq:equilibrium}
\end{equation}
Thus the distance process exhibits a linear mean-reverting drift
toward the stochastic equilibrium point.
This structure is analogous to the restoring drift appearing in Ornstein--Uhlenbeck processes (e.g., \cite{Oksendal2003}).
Under appropriate scaling, this structure 
suggests a connection to Ornstein--Uhlenbeck type behavior.  

For completeness, the conditional variance of the distance increment
can also be computed explicitly. Since $\Delta d\in\{+1,0,-1\}$,
\begin{equation}
\mathbb{E}[(\Delta d)^2\mid d]
=
p_d + q_d
=
\frac{(k-d)(n-k-d)+d^2}{k(n-k)}.
\end{equation}
The conditional variance therefore becomes
\begin{equation}
\mathrm{Var}(\Delta d\mid d)
=
\frac{(k-d)(n-k-d)+d^2}{k(n-k)}
-
\left(
1-\frac{nd}{k(n-k)}
\right)^2 .
\end{equation}

\begin{remark}[Entropy-Driven Drift vanishing point vs. Entropy-maximizing point] 
The results derived so far exhibit that 
the entropy-driven drift vanishing point $d^*$ given by Eq.~(\ref{eq:dstar_drift})
is not identical to the entropy-maximizing point $\hat{d}^*$ given by Eq.~(\ref{eq:dstar}). 
The vanishing of the drift is determined by the condition $p_d = q_d$, 
where $p_d$ and $q_d$ denote the transition probabilities of the induced birth-and-death chain. 
This condition is not identical to the maximization of the entropy $S(d)$, 
which is defined through the shell multiplicities. 
However, in the thermodynamic limit 
where $n \rightarrow \infty$ with $k/n = \alpha$ fixed, 
both points converge to the same normalized concentration 
point $\alpha(1-\alpha)$.
\end{remark}

The entropy increment and the entropy-driven drift are closely related but conceptually distinct: 
the former is a static property of shell multiplicities, 
whereas the latter arises dynamically from the asymmetry of local transitions between neighboring shells 
induced by these multiplicities, as quantified by Eq.~(\ref{eq:pd-qd}).

\subsection{Expected Hitting Time to a Target Configuration}

In this subsection, we consider the uniform random walk $\{ X_t \}$ on the Johnson graph $J(n,k)$ 
and analyze the expected hitting time to a fixed target configuration $x^*$.

Due to the distance-transitivity of the Johnson graph (Property 1), 
the expected hitting time to $x^*$ from any configuration $x \in \Omega$ depends only on its initial distance $m = v(X_0, x^*)=Y_0$. 
Formally, for all $x \in \Omega_m$, $\mathbb{E}[\tau_{x^*} \mid X_0 = x] = \mathbb{E}[\tau_0 \mid Y_0 = m]$, 
where $\tau_{x^*} = \inf \{ t \ge 0 : X_t = x^* \}$ is the hitting time to $x^*$ 
and $\tau_0 = \inf \{ t \ge 0 : Y_t = 0 \}$ is the hitting time to distance zero in the distance process.
This allows us to analyze the hitting time of the original search process $\{ X_t \}$ directly 
through the induced distance process $\{ Y_t \}$.
\begin{theorem}[Expected Hitting Time of the Search Process]
Let $\{ X_t \}_{t=0}^\infty$ be the random walk defined in Section~6.1,
and let $x^* \in \Omega$ be a fixed target configuration.
For any initial configuration $X_0 \in \Omega$ with $v(X_0, x^*) = m$ $(m=0,1,\ldots,d_{\max})$,  
the expected hitting time 
satisfies:
\begin{equation}
\mathbb{E}[\tau_{x^*} \mid v(X_0, x^*) = m] = \sum_{i=1}^{m} \frac{1}{q_i |\Omega_i|} \sum_{j=i}^{d_{\max}} | \Omega_j|,
\label{eq:hitting_time_general}
\end{equation}
where $|\Omega_j|$ denotes the cardinality of the distance shell $\Omega_j$, 
which is given by Eq.~(\ref{eq:shell_size}), 
and $q_i$ is the downward transition probability in the distance process $\{ Y_t \}$, 
which is given by Eq.~(\ref{eq:qd}). 
\end{theorem}
\begin{proof}
First, 
we establish that the analysis of the search process $\{ X_t \}$ can be reduced to 
the distance process $\{ Y_t \}$ on $\{0, \dots, d_{\max}\}$. 
For the search process $\{ X_t \}$, 
the expected hitting time from $x$ to $x^*$ 
depends only on the distance $m = v(x, x^*)$ 
due to the distance-transitivity of $J(n,k)$ (Property~1), 
which ensures that all states at the same distance are equivalent under the dynamics. 
Thus, we define $h_m = \mathbb{E}[\tau_{x^*} \mid X_0 \in \Omega_m] = \mathbb{E}[\tau_0 \mid Y_0 = m]$
for $m \in \{ 0,\ldots, d_{\max} \}$. 

For the birth-and-death chain $\{ Y_t \}$, from its Markov property, 
the first-step analysis for $i \in \{1, \dots, d_{\max}-1\}$ yields 
\begin{equation}
h_i = 1 + p_i h_{i+1} + q_i h_{i-1} + r_i h_i, 
\label{eq:first-step}
\end{equation}
where $p_i$, $q_i$, and $r_i$ are the transition probabilities defined in Eqs.~(\ref{eq:pd}), (\ref{eq:qd}), and (\ref{eq:rd}), respectively.
Let $\delta_i = h_i - h_{i-1}$ for $i \in \{1,\dots,d_{\max}\}$.
For $i \in \{1,\dots,d_{\max}-1\}$, adding $p_i h_i + q_i h_i$ to both sides of Eq.~(\ref{eq:first-step}), 
using $p_i + q_i + r_i = 1$, and rearranging gives
\begin{equation}
q_i \delta_i = 1 + p_i \delta_{i+1}.
\label{eq:recurrence}
\end{equation}
At the boundary $i=d_{\max}$, since $p_{d_{\max}}=0$, the first-step equation reduces to
\begin{equation}
q_{d_{\max}} \delta_{d_{\max}} = 1.
\label{eq:boundary_recurrence}
\end{equation}
Next, from Eqs.~(\ref{eq:qd}), (\ref{eq:pd}), and (\ref{eq:shell_size}), we have
\begin{equation}
\frac{p_i}{q_{i+1}}
=
\frac{(k-i)(n-k-i)}{(i+1)^2}
=
\frac{|\Omega_{i+1}|}{|\Omega_i|},
\label{eq:ratio_transition_probabilities}
\end{equation}
which implies the detailed balance relation (see, e.g., \cite{LevinPeresWilmer2009})
\begin{equation}
|\Omega_i| p_i = |\Omega_{i+1}| q_{i+1}.
\label{eq:detailed_balance}
\end{equation}
Multiplying Eq.~(\ref{eq:recurrence}) by $|\Omega_i|$ and using Eq.~(\ref{eq:detailed_balance}) yields, for $i \in \{1,\dots,d_{\max}-1\}$,
\begin{equation}
|\Omega_i| q_i \delta_i
=
|\Omega_i|
+
|\Omega_{i+1}| q_{i+1} \delta_{i+1}.
\label{eq:recurrence_2}
\end{equation}
Define
\begin{equation}
A_i = |\Omega_i| q_i \delta_i.
\label{eq:Ai}
\end{equation}
Then the above relation (\ref{eq:recurrence_2}) becomes
\begin{equation}
A_i = |\Omega_i| + A_{i+1},
\qquad i=1,\dots,d_{\max}-1,
\label{eq:recurrence_Ai}
\end{equation}
while the boundary condition (\ref{eq:boundary_recurrence}) gives
\begin{equation}
A_{d_{\max}} = |\Omega_{d_{\max}}|.
\label{eq:boundary_Ai}
\end{equation}
By backward iteration of Eq.~(\ref{eq:recurrence_Ai}) with the boundary condition (\ref{eq:boundary_Ai}), we obtain
\begin{equation}
A_i = \sum_{j=i}^{d_{\max}} |\Omega_j|,
\label{eq:Ai_solution}
\end{equation}
and therefore, from Eqs.~(\ref{eq:Ai}) and (\ref{eq:Ai_solution}), we have  
\begin{equation}
\delta_i
=
\frac{1}{q_i |\Omega_i|}
\sum_{j=i}^{d_{\max}} |\Omega_j|.
\label{eq:delta_i_solution}
\end{equation}
Finally, since $h_0=0$ and $h_m=\sum_{i=1}^m \delta_i$, from Eq.~(\ref{eq:delta_i_solution}),  we conclude that
\begin{equation}
h_m
=
\sum_{i=1}^{m} \frac{1}{q_i |\Omega_i|} \sum_{j=i}^{d_{\max}} |\Omega_j|, 
\end{equation}
which completes the proof.
\end{proof}

In the fixed-ratio regime where $k/n=\alpha$ with $0<\alpha<1$,
the expected hitting time grows exponentially in $n$.
This follows from the representation (\ref{eq:hitting_time_general})
together with the exponential growth of the shell cardinalities,
whose dominant contribution comes from shells near the
entropy-maximizing distance.
Using the standard entropy approximation of binomial coefficients
\cite{CoverThomas2006} and asymptotic properties of random walks
on combinatorial structures \cite{LevinPeresWilmer2009}, 
the leading exponential scale is governed by 
\(H(\alpha)=-\alpha\log\alpha-(1-\alpha)\log(1-\alpha)\).

\subsection{Idealized Objective Guidance}

We now extend the baseline analysis to incorporate
objective-driven effects.
To this end, we consider a Metropolis dynamics \cite{LevinPeresWilmer2009} on the Johnson graph
with an idealized objective function
\begin{equation}
f(x) = v(x,x^*),
\end{equation}
where $x^*$ is the reference configuration.
This setting assumes that the distance to the target is directly observable,
thereby removing informational limitations and allowing us to isolate
the interaction between objective-driven and entropy-driven components.

Under this dynamics, candidate moves are generated according to the same
random swap mechanism as in the uniform random walk.
A proposed move from $x$ to $x'$ is accepted with probability
\begin{equation}
\min\left(1, \exp(-\beta (f(x') - f(x)))\right),
\end{equation}
where $\beta \ge 0$ controls the strength of the objective guidance.
When $\beta=0$, the process reduces to the baseline random walk,
while larger values of $\beta$ increasingly favor moves toward the target.

Since the objective depends only on the distance,
the induced distance process $\{Y_t\}$ remains a birth-and-death chain.
In particular, inward transitions (decreasing distance) are always accepted,
while outward transitions are accepted with probability $e^{-\beta}$.
Therefore, the drift of the distance process is given by
\begin{equation}
\mathbb{E}[\Delta d \mid d]
=
p_d e^{-\beta} - q_d,
\label{eq:metropolis_drift}
\end{equation}
where $p_d$ and $q_d$ are the transition probabilities
defined in Eqs.~(\ref{eq:pd}) and (\ref{eq:qd}).
Using their explicit expressions, we obtain
\begin{equation}
\mathbb{E}[\Delta d \mid d]
=
\frac{(k-d)(n-k-d)e^{-\beta} - d^2}{k(n-k)}.
\label{eq:metropolis_drift_explicit}
\end{equation}
This expression (\ref{eq:metropolis_drift_explicit}) shows how objective-based reweighting modifies
the baseline entropy-driven drift by suppressing outward transitions.
For finite $\beta$, outward transitions remain possible, and the
structural contribution associated with local transition asymmetry
therefore remains present in the dynamics.

For $\beta > 0$, the distance $d^*$ at which the drift vanishes is given by
\begin{equation}
d^* =
\frac{-n + \sqrt{n^2 + 4(e^\beta - 1)k(n-k)}}
{2(e^\beta - 1)}.
\label{eq:metropolis_drift_vanishing_point}
\end{equation}
The case $\beta=0$ is recovered by taking the limit of Eq.~(\ref{eq:metropolis_drift_vanishing_point}) as $\beta \to 0$, which yields
$d^*=k(n-k)/n$.
As $\beta$ increases, $d^*$ decreases,
reflecting the increasing influence of objective guidance.
However, for any finite $\beta$, the equilibrium remains separated
from the target.

The expected hitting time under this dynamics can be derived
by modifying Theorem~1 to account for the reweighting of outward transitions. 
The Metropolis reweighting changes the reversible weights of the
distance process from the purely structural weights \(|\Omega_i|\)
to weights proportional to \(|\Omega_i|e^{-\beta i}\).
Applying the standard hitting-time formula for birth-and-death chains
with these weights and simplifying gives 
\begin{equation}
\mathbb{E}_\beta[\tau_{x^*} \mid v(X_0,x^*) = m]
=
\sum_{i=1}^{m} \frac{1}{q_i |\Omega_i|}
\sum_{j=i}^{d_{\max}} |\Omega_j| e^{-\beta (j-i)}.
\label{eq:hitting_time_metropolis}
\end{equation}
When $\beta=0$, Eq.~(\ref{eq:hitting_time_metropolis}) reduces to the baseline hitting time.
Equation~(\ref{eq:hitting_time_metropolis}) shows that objective-based
reweighting suppresses the contribution of outward transitions,
but does not completely remove it for finite $\beta$.

In the limit $\beta \to \infty$, outward transitions are suppressed,
and the dynamics reduce to a greedy process that only accepts
moves toward the target.
In this regime, the entropy-driven outward bias no longer appears
in the observed dynamics.
However, such dynamics may introduce distinct forms of difficulty
in general optimization problems, such as trapping in local minima.

The drift formula in Eq.~(\ref{eq:metropolis_drift_explicit})
and the hitting-time formula in Eq.~(\ref{eq:hitting_time_metropolis})
show that objective-driven reweighting can mitigate
entropy-driven effects, while the structural origin of the drift
remains present for finite $\beta$.
Thus, the observed dynamics reflect the interaction between
objective-driven and entropy-driven components.

\section{Experimental Illustration of Entropy-Driven Drift}

The theoretical analysis in the previous sections shows that
the structure of the underlying graph induces an asymmetry in
local transitions, which in turn generates a systematic drift
in the distance process toward an equilibrium distance.

The purpose of this section is not to establish new phenomena,
but to provide numerical illustrations of the theoretical results.
In particular, we visualize the trajectory-level behavior implied
by the drift structure derived in Section~6 and confirm how the
induced transition asymmetry shapes the evolution of trajectories.

Throughout the experiments we consider the Johnson graph $J(n,k)$
with $n=200$ and $k=40$.

\subsection{Sample Trajectories of the Distance Process}

\begin{figure}[t]
\centering
\includegraphics[width=\linewidth]{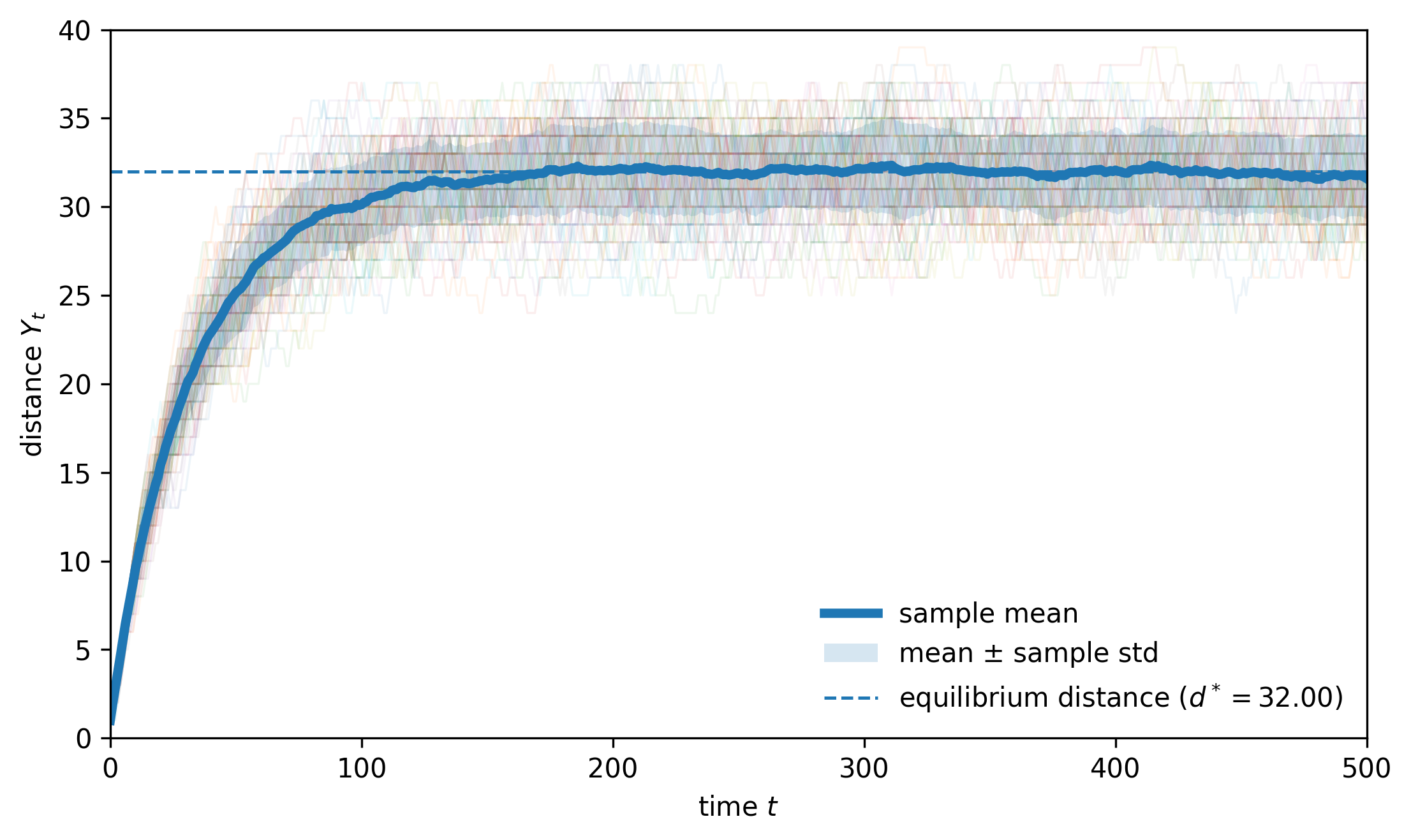}
\caption{
Sample trajectories of the distance process under the uniform random walk on
the Johnson graph $J(200,40)$.
Each trajectory starts from the initial state $d=1$ and evolves
according to random swap moves.
The dashed horizontal line indicates the equilibrium distance $d^* = 32.00$ where the drift vanishes.
The mean trajectory (thick curve) approaches this
equilibrium distance, illustrating the mean-reverting behavior
predicted by the drift analysis.
}
\label{fig:distance_trajectories}
\end{figure}

To illustrate the drift structure analyzed in Section~6,
we examine the behavior of the distance process through simulation.

Figure~\ref{fig:distance_trajectories} shows 100 sample trajectories
of the distance process.
Each trajectory starts from the initial state $d=1$, which is close
to the target configuration, and evolves according to the random
walk on the Johnson graph $J(200,40)$.
The thick curve in the figure represents the mean trajectory,
while the shaded area indicates one standard deviation around the mean.
The dashed horizontal line indicates the equilibrium distance
$d^* = 32.00$, as given by Eq.~(\ref{eq:dstar_drift}) for $n=200$ and $k=40$.

The trajectories exhibit a systematic drift toward the equilibrium
distance $d^*$.
In particular, even when the initial state is close to the target,
typical trajectories move away from it and concentrate around
$d^*$.
The mean trajectory also approaches this equilibrium distance,
consistent with the mean-reverting drift derived in Section~6.

These results directly illustrate the mechanism identified in the theoretical analysis:
the induced asymmetry of local transitions generates a drift
that biases trajectories toward the equilibrium distance,
as predicted by the analytical expression of the drift.

\subsection{Entropy Landscape and Drift Structure}

\begin{figure}[t]
\centering
\includegraphics[width=\linewidth]{figures/entropy_landscape.png}
\caption{
Entropy landscape, its increment, and the induced drift
in the random walk on the Johnson graph $J(200,40)$.
The left panel shows the entropy landscape $S(d)$,
the middle panel shows the entropy increment $\Delta S(d)$,
and the right panel shows the drift $\mathbb{E}[\Delta d \mid d]$ of the distance process.
The dashed vertical line indicates the equilibrium distance $d^* = 32.00$.
}
\label{fig:entropy_landscape}
\end{figure}

We next illustrate the structural origin of the trajectory behavior
shown in Fig.~\ref{fig:distance_trajectories}.

Figure~\ref{fig:entropy_landscape} shows the entropy landscape
$S(d)$ together with the corresponding entropy increment
$\Delta S(d)$ and the drift $\mathbb{E}[\Delta d \mid d]$
as a function of $d$, as given by
Eqs.~(\ref{eq:entropy-landscape}), (\ref{eq:entropy-increment}),
and (\ref{eq:drift}), respectively.
The dashed vertical line indicates the equilibrium distance
$d^* = 32.00$, where the drift vanishes.

The entropy landscape exhibits a unimodal structure with its
maximum located away from the target state.
For the present parameters, the continuous maximizer is
$\hat{d}^* = 31.68$, and the discrete maximum occurs at the
nearest shell(s).
This confirms that the combinatorial mass of the state space is
concentrated at intermediate distances.

The entropy increment $\Delta S(d)$ is positive for
$d < \hat{d}^*$ and negative for $d > \hat{d}^*$,
indicating that the entropy increases as the distance approaches
the entropy-maximizing region.
Near the target, the entropy increment $\Delta S(d)$ becomes strongly positive,
reflecting a sharp imbalance in the number of neighboring states.

The drift $\mathbb{E}[\Delta d \mid d]$ is positive for
$d < d^*$ and negative for $d > d^*$,
showing that the dynamics exhibit a systematic tendency toward
the equilibrium distance.
This behavior is consistent with the mean-reverting drift
derived in Eq.~(\ref{eq:equilibrium}).

These observations illustrate the mechanism identified in the
theoretical analysis.
The combinatorial distribution of states determines the entropy
landscape, while the local transition structure determines how
this distribution is translated into a directional bias in the
dynamics.
In particular, the asymmetry in the number of edges connecting
neighboring shells induces an imbalance between outward and
inward transitions, which gives rise to the drift observed in
the distance process.

\subsection{Idealized Objective Guidance and Entropy Competition}

\begin{figure}[t]
\centering
\includegraphics[width=\linewidth]{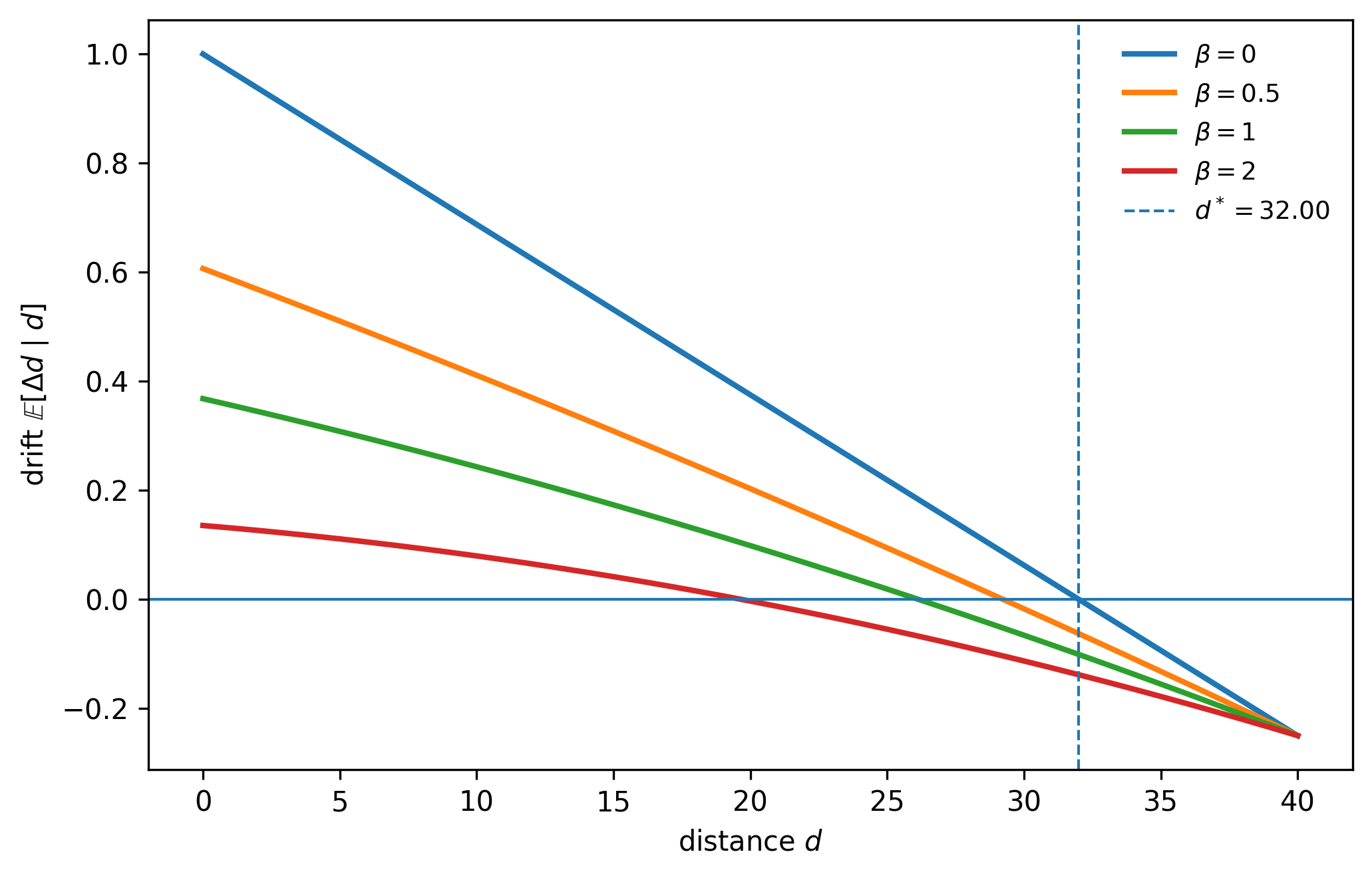}
\caption{
Interaction between idealized objective guidance and entropy-driven drift.
The figure shows the drift of the distance process 
under Metropolis dynamics for various values of $\beta$. 
The parameter $\beta$ represents the strength of objective-based guidance toward the target. 
When $\beta=0$ (uniform random walk), the drift is dominated by entropy-driven drift across the entire space. 
As $\beta$ increases, the downward shift of the curves illustrates the effect of objective guidance. 
Crucially, even for $\beta=2$, the drift can remain positive in the vicinity of the target, 
showing that, for moderate levels of objective guidance, entropy-driven drift can remain dominant near the target.
}
\label{fig:objective_drift}
\end{figure}

To examine the interaction between structural drift and
objective-based guidance, we consider the behavior of the
distance process under Metropolis dynamics considered in Section~6.5, 
with varying values of $\beta$.

Figure~\ref{fig:objective_drift} shows the drift of the distance process.
As $\beta$ increases, the drift curve shifts downward,
reflecting the increasing influence of objective-based transitions.
However, for moderate values of $\beta$, the drift remains positive
in the vicinity of the target, indicating that the outward bias
induced by the transition structure persists.

\begin{figure}[t]
\centering
\includegraphics[width=\linewidth]{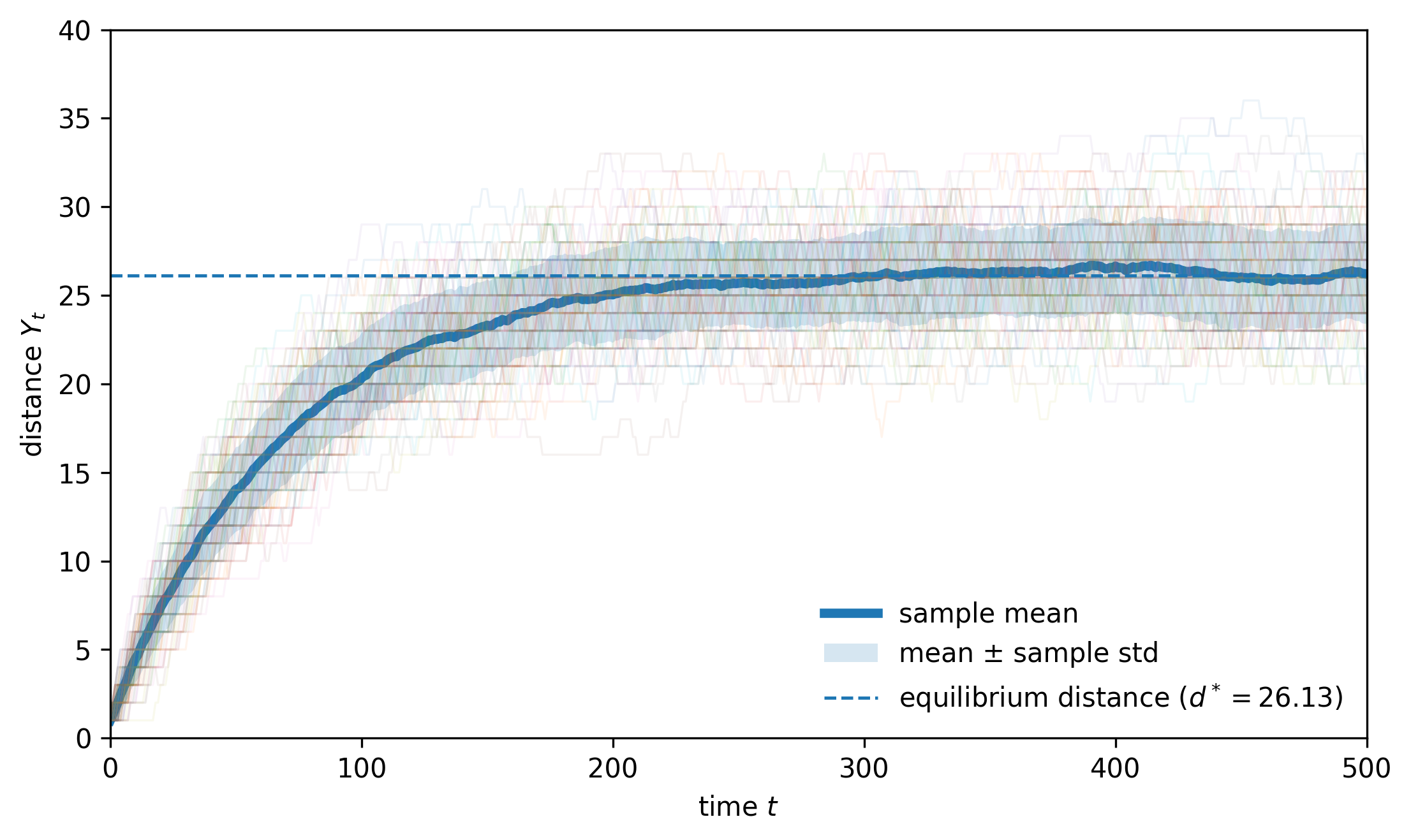}
\caption{
Sample trajectories of the distance process under idealized
objective guidance ($\beta=1$), starting from $d=1$.
While the objective introduces a tendency toward the target,
the trajectories do not concentrate at $d=0$, but instead
stabilize around a nonzero equilibrium distance $d^*=26.13$.
This behavior reflects the competition between objective-driven
drift toward the target and entropy-driven drift away from it.
}
\label{fig:distance_trajectories_beta}
\end{figure}

Figure~\ref{fig:distance_trajectories_beta} shows sample trajectories
starting near the target.
Despite the presence of objective guidance, trajectories do not
concentrate at $d=0$, but instead stabilize around a nonzero distance.
This behavior reflects the competition between inward
objective-driven transitions and outward transitions induced by
the underlying graph structure.

\begin{figure}[t]
\centering
\includegraphics[width=\linewidth]{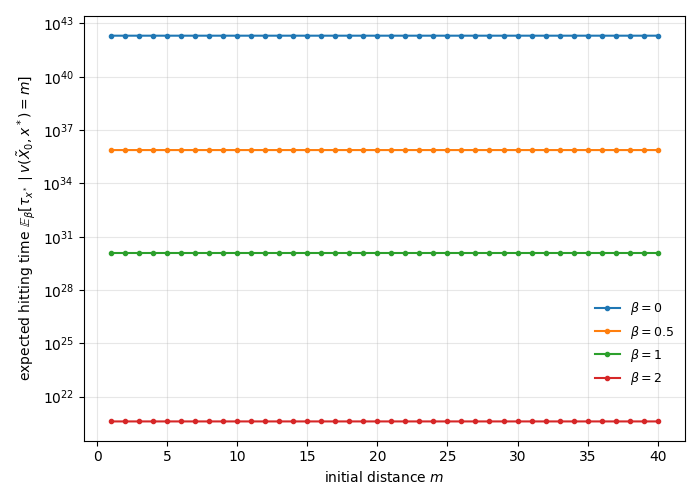}
\caption{
Expected hitting time to the target configuration under the idealized objective guidance $f(x)=v(x,x^*)$, 
shown as a function of the initial distance $m$. 
Increasing $\beta$ reweights the contribution of outward trajectories,
leading to a substantial reduction in expected hitting time across all initial distances.
}
\label{fig:expected_hitting_time_MH}
\end{figure}

Figure~\ref{fig:expected_hitting_time_MH} shows the expected hitting time
as a function of the initial distance.
Increasing $\beta$ reduces the expected hitting time,
while the dependence on the initial distance remains comparatively weak.
This indicates that the reweighting of local transitions,
rather than the starting position, governs the behavior of the process.

These observations directly illustrate the mechanism identified in the theoretical analysis: 
objective-based reweighting modifies the balance between inward
and outward transitions, but for finite $\beta$ it does not alter
the underlying transition structure.
As a result, the induced directional bias in the dynamics remains present,
although its effect on the trajectory distribution can be attenuated.

\subsection{Comparison with IID Uniform Sampling}

\begin{figure}[t]
\centering
\includegraphics[width=\linewidth]{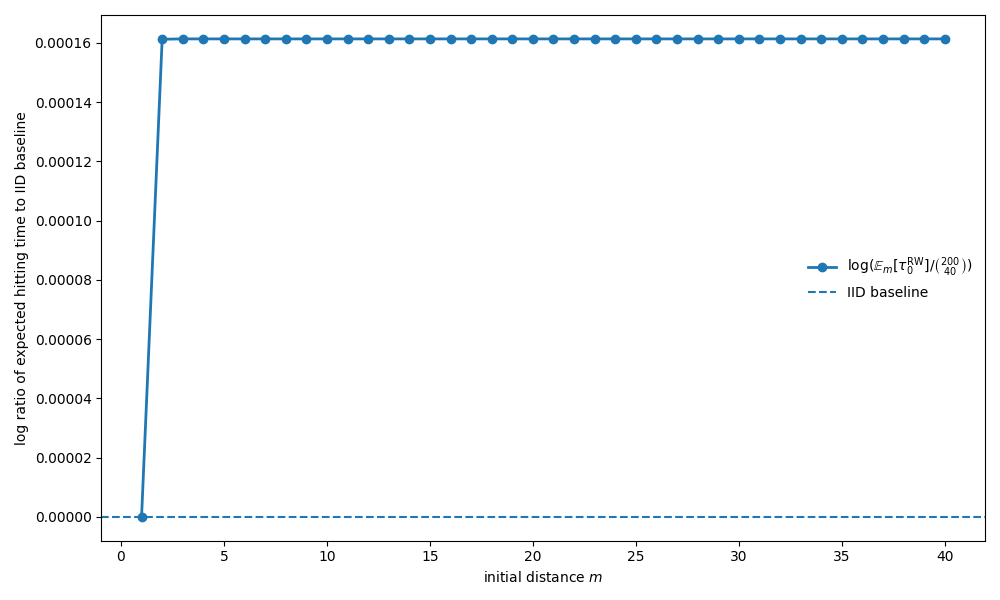}
\caption{
Logarithmic ratio of expected hitting time of the random walk relative to IID uniform sampling. 
The vertical axis shows
$\log\!\bigl(\mathbb{E}_m[\tau_0^{\mathrm{RW}}] / \binom{n}{k}\bigr)$,
where $\binom{n}{k}$ is the IID baseline corresponding to the combinatorial rarity of the target.
Positive values indicate that the random walk requires longer search times than IID sampling,
reflecting the effect of locality in the transition dynamics.
}
\label{fig:hitting_time_ratio}
\end{figure}

In this subsection, to clarify the role of locality, we compare the random walk on the
Johnson graph with IID (independent and identically distributed) uniform sampling over the state space.

Under IID uniform sampling, each step selects a configuration
uniformly from all $k$-subsets.
Since the probability of hitting the target at each step is
$1/\binom{n}{k}$, the expected hitting time is
$\mathbb{E}[\tau_0^{\mathrm{IID}}] = \binom{n}{k}$.
This baseline reflects only the combinatorial rarity of the target
and involves no constraints on transitions.

By contrast, the random walk is constrained by local transitions.
As shown in Section~6, this locality induces a directional bias
in the distance process, so that the hitting time depends not only on
the size of the state space but also on the transition structure.

To quantify the additional cost induced by locality, we consider the ratio of the expected hitting time
of the random walk to the IID baseline, which is given by
\begin{equation}
\log\!\left(\frac{\mathbb{E}_m[\tau_0^{\mathrm{RW}}]}{\binom{n}{k}}\right),
\end{equation}
where $\mathbb{E}_m[\tau_0^{\mathrm{RW}}]$ is the expected hitting
time of the random walk starting from distance $m$, 
which is computed from Eq.~(\ref{eq:hitting_time_general}).

Figure~\ref{fig:hitting_time_ratio} shows that the random walk requires
longer expected hitting times than the IID baseline,
except for the special case $m=1$, where the random walk can reach the target in a single step,
resulting in $\mathbb{E}_{1}[\tau_0^{\mathrm{RW}}] = \binom{n}{k} - 1$.
This reflects the fact that, at distance $m=1$, the random walk can reach the target in a single step. 
Moreover,
the ratio is largely insensitive to the initial distance $m$
for $m \ge 2$.
This observation indicates that, once the process is separated from the target,
the dominant factor governing the hitting time is not geometric proximity,
but the transition-induced bias in the trajectory dynamics.

Notably, both processes share the same stationary distribution,
yet exhibit fundamentally different hitting-time behavior.
This shows that optimization difficulty is not determined by the
rarity of the target alone.
Rather, locality imposes structural constraints on transitions,
leading to trajectory behavior that differs fundamentally
from that of IID sampling and results in an additional hitting-time cost
beyond the combinatorial baseline.

This comparison highlights that state-based properties such as
multiplicity or stationary distributions are insufficient to
characterize search dynamics.
Instead, the transition kernel governing the process plays a
central role in shaping trajectory-level behavior.

\section{Discussion}

The analysis in this paper clarifies how a static combinatorial
structure can give rise to a dynamical bias in trajectory evolution.
The key point is that shell multiplicities alone do not constitute
a drift. Rather, drift arises only when these multiplicities are coupled
with local transition constraints on the underlying graph.

In the Johnson graph, this coupling can be analyzed explicitly.
The distance-transitivity of the graph implies that the distance
process is Markovian, and the imbalance between inward and outward
local transitions can be expressed solely as a function of the
distance from the reference configuration. This makes it possible
to identify the structural component of the drift induced by the
underlying graph.

This perspective on drift induced by local transition asymmetry 
also distinguishes the entropy landscape from the
entropy-driven drift. The entropy landscape describes how the volume
of the state space is distributed across distance shells. 
By contrast,
the entropy-driven drift is a property of the transition dynamics:
it describes how local moves convert this static imbalance
into a systematic bias in trajectory evolution.

The comparison with objective-guided dynamics further shows how
the structural and objective-driven components interact. Reweighting
local transitions through an objective function can counteract the
structural drift, but for finite reweighting the underlying graph
structure remains unchanged. Thus the observed dynamics reflect the
combined effect of objective-based preferences and structural
transition asymmetry. 
From this viewpoint, objective guidance can be interpreted as
modifying the effective weights over distance shells,
while leaving the local transition structure intact, 
thereby altering the balance between inward and outward moves. 

While entropy has long been recognized as a key factor
in shaping behavior in combinatorial systems,
it is often reflected through distributions or equilibrium properties.
The present work highlights a complementary viewpoint,
in which entropy contributes to dynamics through
asymmetry in local transitions,
thereby inducing a bias in trajectory evolution.
This perspective suggests that optimization difficulty may arise
not only from the landscape of states,
but also from the geometry of trajectories induced by local transitions.

The Johnson graph provides a highly symmetric setting in which
the distance to the target offers a natural coordinate for
describing the dynamics.
In this case, the entropy-driven drift can be expressed explicitly
as a function of the distance, and the resulting behavior is
well captured by a one-dimensional reduction.
In more general combinatorial state spaces, however, such symmetry
need not be present.
The relationship between distance and transition structure may
then become more complex, and the induced drift may no longer be
fully characterized by the distance alone.
In particular, trajectories that do not follow shortest paths
with respect to the target may become dynamically favored under the induced
dynamics, depending on the structure of local transitions.
Understanding how the underlying graph structure shapes the
effective trajectory geometry beyond shortest-path behavior,
and how this influences optimization performance,
remains an important direction for future work.

\section{Conclusion}

This paper analyzes how the structure of a combinatorial state space,
together with local transition rules, shapes search dynamics.
Using the Johnson graph as a canonical setting, we show that
asymmetry in local transitions induces a systematic bias in the
distance process under the uniform random walk.

This bias arises from the imbalance between inward and outward
transitions and can be expressed explicitly as a function of the
distance from a reference configuration.
The resulting dynamics exhibit mean-reverting behavior toward
an equilibrium distance determined by the transition structure.

A key contribution of this work is to distinguish the entropy
landscape from the induced drift.
While the entropy landscape describes the distribution of states
across distance shells, the drift is a dynamical property that
arises from the interaction between shell multiplicities and
local transition constraints.

The explicit characterization obtained in the Johnson graph relies
on its symmetry properties, which allow the dynamics to be reduced
to a one-dimensional birth-and-death process.
For more general graphs, such a reduction need not hold, and the
drift may depend on the detailed structure of the underlying graph.

These results suggest that optimization difficulty may arise not only
from properties of individual states, but also from the geometry of
trajectories induced by local transitions.
Understanding this trajectory-level structure in more general settings
remains an important direction for future work.

\bibliographystyle{plain} 
\bibliography{refs}

@book{HoosStutzle2005,
  author    = {Holger H. Hoos and Thomas St{\"u}tzle},
  title     = {Stochastic Local Search: Foundations and Applications},
  publisher = {Elsevier / Morgan Kaufmann},
  year      = {2004},
  isbn      = {978-1558608726}
}

@book{NeumannWitt2010,
  author    = {Frank Neumann and Carsten Witt},
  title     = {Bioinspired Computation in Combinatorial Optimization: Algorithms and Their Computational Complexity},
  series    = {Natural Computing Series},
  publisher = {Springer},
  year      = {2010},
  doi       = {10.1007/978-3-642-16544-6},
  isbn      = {978-3642165439}
}

@book{RobertCasella2004,
  title={Monte Carlo Statistical Methods},
  author={Robert, Christian P. and Casella, George},
  year={2004},
  publisher={Springer}
}

@book{DoerrNeumann2020,
  editor    = {Benjamin Doerr and Frank Neumann},
  title     = {Theory of Evolutionary Computation: Recent Developments in Discrete Optimization},
  publisher = {Springer},
  year      = {2020},
  doi       = {10.1007/978-3-030-29414-4}
}

@book{AugerDoerr2011,
  title={Theory of Randomized Search Heuristics},
  author={Auger, Anne and Doerr, Benjamin},
  year={2011},
  publisher={World Scientific}
}

@phdthesis{Jones1995,
  author    = {Terry Jones},
  title     = {Evolutionary Algorithms, Fitness Landscapes and Search},
  school    = {University of New Mexico},
  year      = {1995},
  address   = {Albuquerque, NM},
}

@book{GodsilRoyle2001,
  author = {Chris Godsil and Gordon Royle},
  title = {Algebraic Graph Theory},
  publisher = {Springer},
  year = {2001}
}

@book{LevinPeresWilmer2009,
  author = {David A. Levin and Yuval Peres and Elizabeth L. Wilmer},
  title = {Markov Chains and Mixing Times},
  publisher = {American Mathematical Society},
  year = {2009}
}

@book{CoverThomas2006,
  author = {Thomas M. Cover and Joy A. Thomas},
  title = {Elements of Information Theory},
  publisher = {Wiley},
  year = {2006}
}

@ARTICLE{Benidis2018,
  author={Benidis, Konstantinos and Feng, Yiyong and Palomar, Daniel P.},
  journal={IEEE Transactions on Signal Processing}, 
  title={Sparse Portfolios for High-Dimensional Financial Index Tracking}, 
  year={2018},
  volume={66},
  number={1},
  pages={155-170},
  doi={10.1109/TSP.2017.2762286}
}

@book{KemenySnell1976,
  author = {John G. Kemeny and J. Laurie Snell},
  title = {Finite Markov Chains},
  publisher = {Springer-Verlag},
  year = {1976}
}

@incollection{Lengler2020,
  author    = {Johannes Lengler},
  title     = {Drift Analysis},
  booktitle = {Theory of Evolutionary Computation: Recent Developments in Discrete Optimization},
  publisher = {Springer},
  year      = {2020},
  pages     = {89--131},
  doi       = {10.1007/978-3-030-29414-4_3}
}

@article{He2001, 
title = {Drift analysis and average time complexity of evolutionary algorithms},
journal = {Artificial Intelligence},
volume = {127},
number = {1},
pages = {57-85},
year = {2001},
issn = {0004-3702},
doi = {https://doi.org/10.1016/S0004-3702(01)00058-3},
url = {https://www.sciencedirect.com/science/article/pii/S0004370201000583},
author = {Jun He and Xin Yao}
}

@article{Malan2013,
  author    = {Katherine M. Malan and Andries P. Engelbrecht},
  title     = {A Survey of Techniques for Characterising Fitness Landscapes and Some Possible Ways Forward},
  journal   = {Information Sciences},
  volume    = {241},
  pages     = {148--163},
  year      = {2013},
  doi      = {10.1016/j.ins.2013.04.015},
  publisher = {Elsevier}
}

@inproceedings{Ochoa2008,
author = {Ochoa, Gabriela and Tomassini, Marco and V\'{e}rel, Seb\'{a}stien and Darabos, Christian},
title = {A study of {NK} landscapes' basins and local optima networks},
year = {2008},
isbn = {9781605581309},
publisher = {Association for Computing Machinery},
address = {New York, NY, USA},
url = {https://doi.org/10.1145/1389095.1389204},
doi = {10.1145/1389095.1389204},
booktitle = {Proceedings of the 10th Annual Conference on Genetic and Evolutionary Computation},
pages = {555--562},
numpages = {8},
location = {Atlanta, GA, USA},
series = {GECCO '08}
}

@Inbook{Ochoa2014,
author="Ochoa, Gabriela
and Verel, S{\'e}bastien
and Daolio, Fabio
and Tomassini, Marco",
editor="Richter, Hendrik
and Engelbrecht, Andries",
title="Local Optima Networks: A New Model of Combinatorial Fitness Landscapes",
bookTitle="Recent Advances in the Theory and Application of Fitness Landscapes",
year="2014",
publisher="Springer Berlin Heidelberg",
address="Berlin, Heidelberg",
pages="233--262",
isbn="978-3-642-41888-4",
doi="10.1007/978-3-642-41888-4\_9",
url="https://doi.org/10.1007/978-3-642-41888-4\_9"
}

@InProceedings{Thomson2024,
author="Thomson, Sarah L.
and Ochoa, Gabriela
and van den Berg, Daan
and Liang, Tianyu
and Weise, Thomas",
editor="Affenzeller, Michael
and Winkler, Stephan M.
and Kononova, Anna V.
and Trautmann, Heike
and Tu{\v{s}}ar, Tea
and Machado, Penousal
and B{\"a}ck, Thomas",
title="Entropy, Search Trajectories, and Explainability for Frequency Fitness Assignment",
booktitle="Parallel Problem Solving from Nature -- PPSN XVIII",
year="2024",
publisher="Springer Nature Switzerland",
address="Cham",
pages="377--392",
isbn="978-3-031-70055-2"
}

@inproceedings{Fieldsend2025,
author = {Fieldsend, Jonathan and Liefooghe, Arnaud and Malan, Katherine and Verel, S\'{e}bastien},
title = {Local Optima Networks for Constrained Search Spaces},
year = {2025},
isbn = {9798400714658},
publisher = {Association for Computing Machinery},
address = {New York, NY, USA},
url = {https://doi.org/10.1145/3712256.3726303},
doi = {10.1145/3712256.3726303},
booktitle = {Proceedings of the Genetic and Evolutionary Computation Conference},
pages = {204--212},
numpages = {9},
location = {NH Malaga Hotel, Malaga, Spain},
series = {GECCO '25}
}

@article{Ochoa2021,
title = {Search trajectory networks: A tool for analysing and visualising the behaviour of metaheuristics},
journal = {Applied Soft Computing},
volume = {109},
pages = {107492},
year = {2021},
issn = {1568-4946},
doi = {https://doi.org/10.1016/j.asoc.2021.107492},
url = {https://www.sciencedirect.com/science/article/pii/S1568494621004154},
author = {Gabriela Ochoa and Katherine M. Malan and Christian Blum}
}

@book{Mezard2009,
  author    = {Marc M{\'e}zard and Andrea Montanari},
  title     = {Information, Physics, and Computation},
  publisher = {Oxford University Press},
  year      = {2009}
}

@book{Nishimori2001,
  author    = {Hidetoshi Nishimori},
  title     = {Statistical Physics of Spin Glasses and Information Processing: An Introduction},
  publisher = {Oxford University Press},
  year      = {2001}
}

@book{Peliti2024,
  author    = {Luca Peliti},
  title     = {Statistical Mechanics in a Nutshell, Second Edition},
  publisher = {Princeton University Press},
  year      = {2024}
}

@book{Oksendal2003,
  author    = {Bernt \O{}ksendal},
  title     = {Stochastic Differential Equations: An Introduction with Applications},
  edition   = {6},
  series    = {Universitext},
  publisher = {Springer},
  address   = {Berlin, Heidelberg},
  year      = {2003},
  doi       = {10.1007/978-3-642-14394-6}
}

@article{Guyon2003,
  author    = {Isabelle Guyon and Andr{\'e} Elisseeff},
  title     = {An Introduction to Variable and Feature Selection},
  journal   = {Journal of Machine Learning Research},
  volume    = {3},
  pages     = {1157--1182},
  year      = {2003}
}

@article{BlumLangley1997,
title = {Selection of relevant features and examples in machine learning},
journal = {Artificial Intelligence},
volume = {97},
number = {1},
pages = {245-271},
year = {1997},
issn = {0004-3702},
doi = {https://doi.org/10.1016/S0004-3702(97)00063-5},
url = {https://www.sciencedirect.com/science/article/pii/S0004370297000635},
author = {Avrim L. Blum and Pat Langley}
}

@article{Sakurai2024,
  author  = {Sakurai, Yutaka and Wakabayashi, Daiki and Ishizaki, Fumio},
  title   = {Formulations to select assets for constructing sparse index tracking portfolios},
  journal = {Journal of Investment Strategies},
  year    = {2024},
  volume  = {13},
  number  = {2},
  pages   = {1--15},
  doi     = {10.21314/JOIS.2024.007},
  url     = {https://www.risk.net/journal-of-investment-strategies/7959857/formulations-to-select-assets-for-constructing-sparse-index-tracking-portfolios}
}

\end{document}